\documentclass[aps,prl,twocolumn,superscriptaddress, longbibliography]{revtex4-2}
\usepackage{pdfpages}
\makeatletter
\AtBeginDocument{\let\LS@rot\@undefined}
\makeatother

\usepackage[utf8]{inputenc}
\usepackage[T1]{fontenc}
\usepackage{microtype}
\usepackage[english]{babel}
\usepackage{amsmath}
\usepackage{newtx}
\usepackage{physics}
\usepackage{nicefrac}
\usepackage{xcolor}
\definecolor{cset-aps-blueberry}{RGB}{28,128,158}
\definecolor{cset-aps-blue}{RGB}{46,44,184}
\definecolor{cset-aps-turquoise}{RGB}{0,67,88}
\definecolor{cset-aps-limegreen}{RGB}{190,219,67}
\definecolor{cset-aps-green}{RGB}{31,138,112}
\definecolor{cset-aps-yellow}{RGB}{255,225,25}
\definecolor{cset-aps-orange}{RGB}{253,116,0}
\definecolor{cset-aps-red}{RGB}{219,0,43}
\definecolor{cset-aps-violett}{RGB}{142,68,173}

\usepackage{hyperref}
\hypersetup{colorlinks=true,
	linkcolor={cset-aps-red},
	linkbordercolor={cset-aps-red},
	filecolor={cset-aps-orange},
	filebordercolor={cset-aps-orange},
	citecolor={cset-aps-blue},
	citebordercolor={cset-aps-blue},
	urlcolor={cset-aps-green},
	urlbordercolor={cset-aps-green},
	menucolor={cset-aps-limegreen},
	menubordercolor={cset-aps-limegreen},
	breaklinks=true,
	pdfborderstyle={/S/U/W 2},
pdfpagemode=UseOutlines,
	pdfstartpage={1},
} 

\usepackage[capitalize]{cleveref}
\usepackage{tikz}

\usetikzlibrary{intersections, calc, external}

\DeclareRobustCommand*\circled[1]{\tikzset{external/export next=false}\tikz[baseline=(char.base)]{
            \node[shape=circle,draw,inner sep=1.25pt, outer sep=0pt] (char) {#1};}}
\begin{document}
\title{Confinement-Induced Resonances in Spherical Shell Traps}

\author{C. Moritz Carmesin}
\email{moritz.carmesin@uni-ulm.de}
\affiliation{Institut f\"ur Quantenphysik and Center for Integrated Quantum Science and Technology (IQST), Universit\"at Ulm, 89081 Ulm, Germany}

\author{Maxim A. Efremov}
\affiliation{Institut f\"ur Quantenphysik and Center for Integrated Quantum Science and Technology (IQST), Universit\"at Ulm, 89081 Ulm, Germany}
\affiliation{German Aerospace Center (DLR), Institute of Quantum Technologies, 89081 Ulm, Germany}

\begin{abstract}
The energy spectrum and corresponding wave functions of two bosonic particles confined in a spherically symmetric shell trap and interacting via a three-dimensional zero-range potential are computed. Confinement-induced resonances, originating entirely from the strong coupling of the relative and center-of-mass motions of the two particles, are identified as avoided crossings for certain values of the shell radius. By working close to the found resonances, these results offer a new way to control the atom-atom interaction in an atomic gas by tuning only the geometrical parameters of the shell.
\end{abstract}

\maketitle

\let\vec\boldsymbol

\paragraph{Introduction.--} 
Ultra-cold atomic bubbles are atomic quantum gases confined to a closed two-dimensional surface like a spherical or ellipsoidal shell. They have recently attracted a lot of attention as the best platform to probe fundamental concepts of few- and many-body quantum theory and statistical mechanics in curved manifolds \cite{Tononi2023,Dubessy_Perrin_2025}. Nowadays, there exist only two working schemes \cite{Carollo.2022, Guo_2022,Jia.2022} for producing atomic shells in labs \cite{Lundblad_2023}. The first is using radio-frequency dressing \cite{ZobayGarraway2001,*ZobayGarraway2004,Garraway_2016,PERRIN2017181} in a microgravity environment \cite{Aveline2020, PENLEY2002691, Zoest2010,LOTZ20181967,CAL}, resulting in an ellipsoidal shell. The second scheme involves an optically confined mixture of two Bose–Einstein condensates and either employs a magic laser wavelength \cite{Jia.2022,Meister_2023}, or combines microgravity \cite{PhysRevA.106.013309} with a Feshbach resonance \cite{RevModPhys.82.1225,TIMMERMANS1999199}, to tune the interspecies interaction. This naturally results in a spherical shell for one of the mixture components. The successful realizations of this novel topology have inspired many groups to re-examine many well-known phenomena that were originally discovered for a flat geometry, but now in the case of quantum atomic bubbles. 
Many studies have explored the influence of curvature on Bose-Einstein condensation \cite{RevModPhys.71.463,PhysRevLett.123.160403,Móller_2020,PhysRevLett.125.010402,PhysRevA.104.063310}, the collective excitations in condensate shells \cite{PhysRevA.75.013611,PhysRevA.98.013609,Padavić_2017,PhysRevA.106.013309}, vortices \cite{RevModPhys.81.647,RevModPhys.82.1301,PhysRevA.102.043305,PhysRevA.103.053306}, the Berezinskii-Kosterlitz-Thouless transition \cite{Kosterlitz_2016,PhysRevResearch.4.013122}, and potential scattering of particles \cite{Zhang_2018, Shi_2017}. 

In these phenomena, the interaction between atoms, leading to non-linear dynamics, plays a crucial role.
However, for a limited number of atoms in a spherical shell, the density of atoms decreases rapidly as $1/r_0^2$ with increasing radius $r_0$ of the shell \cite{Boegel_2023}.   
Therefore, for large thin shells, where we have an ideal curved manifold, the atom-atom interaction becomes negligible. Thus, to have the possibility to experimentally study any interaction-driven phenomenon on shells, one has to find a scheme to adjust the atom-atom interaction. Most importantly, such a scheme cannot be based on magnetic or optical fields, as they are already used to create a shell-shaped trap or to tune the repulsive inter-species interaction.

To address this problem, we propose and study in this Letter confinement-induced resonances (CIRs) as a reliable tool to tune the interaction between tightly confined atoms \cite{CIR}. The use of CIRs provides an effective solution to a critical problem in the investigation of how curvature affects many physical phenomena, occurring in few- and many-body physics, statistical mechanics, and physical kinetics. This is due to the fact that in the presence of a confinement, the scattering properties \cite{Olshanii1998,PhysRevLett.91.163201, Petrov2000} and the binding spectrum \cite{Busch1998} of two atoms show remarkable changes compared to the confinement-free atoms. Here we show that in a spherically symmetric bubble many CIRs appear at certain shell radii $r_0$. To do this, we exactly compute the energy spectrum and the corresponding wave functions of two bosonic particles, which are confined in a \emph{shell-shaped trap} and interact with each other via a \emph{three-dimensional} $s$-wave zero-range potential characterized by the scattering length $a_0$. On this basis, we determine the position and width of the CIRs as a function of $a_0$ and $r_0$.

The found CIRs are identified as avoided crossings between a bound (molecular) state with excitation of the center-of-mass (CoM) motion and a trap (non-molecular) state without CoM excitation, as well as between two trap states. These resonances originate entirely from the strong coupling between the relative and CoM degrees of freedom of the two confined particles. Similar CIRs occur in a system of two identical atoms in an anharmonic trap \cite{PhysRevLett.109.073201,PhysRevA.94.022713}, or with heteronucler atoms in a harmonic confinement \cite{Peano_2005,Saenz2007,Melezhik_2009}. Consequently, a slow change of the shell radius $r_0$ around a bound-trap CIR results in the formation of a two-particle molecule, mediated by the confinement and without requiring a third particle. This is due to the fact that only the total energy of two trapped particles is fixed and can be redistributed between the energies of the relative and the CoM degrees of freedom, as well as their coupling energy. Moreover, the confinement-supported formation of molecules is a basic loss mechanism and provides an experimental scheme to observe the found bound-trap CIRs and thus to explore the stability of the atomic bubble. In addition, the found trap-trap CIRs are of great importance for accurately modeling the effective atom-atom interaction and the corresponding non-linear term in the Gross-Pitaevskii equation, while performing a dimensional reduction of the BEC dynamics from three- to two-dimensional curved manifolds \cite{Moller_2020}.

\paragraph{Confinement-Induced Resonances.--}

The emergence of CIRs can be easily understood in terms of configurations. This term refers here to the spatial geometry of the two-particle wave function, induced not only by the interaction and confinement, but also, and most importantly, by the distribution of energy among the different degrees of freedom. This becomes visible as a varying number of nodes in each degree of freedom. 

\begin{figure}
    \centering
\includegraphics{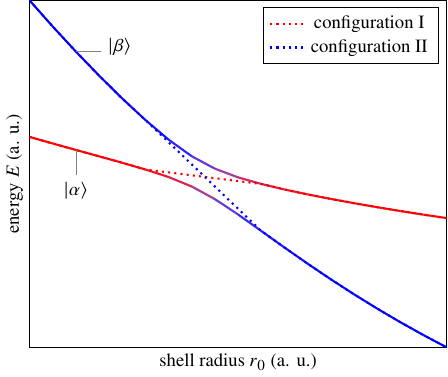}
    \caption{Energy level scheme at a CIR resulting from the coupling of two diabatic states (dotted lines), pertaining to the different configurations I (red) and II (blue). The resulting adiabatic states $\ket{\alpha}$ and $\ket{\beta}$ (solid lines) avoid crossing and allow the adiabatic transition from one configuration into the other by tuning the shell radius. The adiabatic states are colored according to their overlap with the diabatic states by adjusting the transparency of each color: full opacity means full overlap while transparency indicates decreased overlap.}
    
\label{fig:CIR-scheme}
\end{figure}

To illustrate this phenomenon, we present in \cref{fig:CIR-scheme} the general picture of a CIR. For a certain interval of the system parameter, in our case it is the shell radius $r_0$, the energy of the adiabatic state $\ket{\alpha}$ comes close to the one of the state $\ket{\beta}$. For smaller $r_0$, $\ket{\alpha}$ represents configuration I (red color) whereas $\ket{\beta}$ the different configuration II (blue). As an example, at a bound-trap CIR, the configuration I has essentially the form of two non-interacting particles, each of them is in the ground state of the trap, whereas the configuration II describes strong localization in the relative degrees of freedom and excitation in the CoM ones. At the avoided crossing, the configurations mix up, associated with a redistribution of the total energy of the state between different coupled degrees of freedom.  For slightly larger values of $r_0$, the adiabatic states have interchanged their configurations, i.~e. $\ket{\alpha}$ now features the configuration II and vice versa. 

Generally, a CIR occurs when (i) the energies of two different configurations come close to each other (or cross), as shown in \cref{fig:CIR-scheme} by dotted lines, and (ii) the confinement leads to a coupling of these configurations. This coupling induces avoided crossings in the energy spectrum of the adiabatic eigenstates, so that the diabatic states, each of which corresponding to one of the involved configurations, interact strongly.

\paragraph{Two Particles in a Spherical Shell Trap.--} We consider two identical bosons of mass $m$, which are trapped in a shell potential \cite{PhysRevA.98.013609} modeled as a spherically symmetric shifted harmonic oscillator $V_0({\vec r}_i)=\frac{1}{2}m\omega^2(\left|{\vec r}_i\right|-r_0)^2$, where $i=1$ or $2$, with the trap frequency $\omega$ and the shift $r_0$ of the potential minimum. The width of the trapping potential is thus characterized by the oscillator length $a_\text{ho}=\sqrt{\hbar/(m\omega)}$. Instead of the commonly used Fermi-Huang three-dimensional pseudopotential we equivalently model the $s$-wave interparticle interaction by the Bethe-Peierls boundary condition \cite{bethe_quantum_1935} for the two-particle wave function $\Psi$, that is
\begin{subequations}
\begin{equation}
 \label{eqn:bc_BP_r12_1}
   \lim_{\vec r_1\to\vec r_2}\left\{\pdv{\left[|\vec r_1-\vec r_2|\Psi(\vec r_1, \vec r_2)\right]}{|\vec r_1-\vec r_2|}+\frac{|\vec r_1-\vec r_2|}{a_0}\Psi(\vec r_1, \vec r_2)\right\} = 0       
\end{equation}
for partial waves with zero relative angular momentum, $\ell=0$, and
\begin{equation}
 \label{eqn:bc_BP_r12_2}
    \lim_{\vec r_1\to\vec r_2}|\vec r_1-\vec r_2|\Psi(\vec r_1, \vec r_2)=0
\end{equation}
\end{subequations}
for partial waves with the relative angular momentum $\ell>0$.

In the case of an ordinary spherically symmetric harmonic confinement, that is the case of $r_0=0$, both the Hamiltonian 
\begin{equation}
    \hat H = -\frac{\hbar^2}{2m}\left(\Delta_{\vec r_1} + \Delta_{\vec r_2}
\right)+ V_0({\vec  r}_1) +V_0({\vec  r}_2) \label{eqn:H}
\end{equation}
of our system and the boundary conditions, Eqs. \eqref{eqn:bc_BP_r12_1} and \eqref{eqn:bc_BP_r12_2}, are separable in terms of the relative $\vec r = \vec r_1 - \vec r_2$ and CoM $\vec R = \frac{1}{2}(\vec r_1+\vec r_2)$ coordinates and the two-particle energy spectrum can be derived analytically \cite{Busch1998}. Here $\Delta_{\vec r}$ denotes the three-dimensional Laplace operator. In this Letter, we however consider the case of $r_0>0$, giving rise to non-separability of the Hamiltonian in terms of the $\vec r$ and $\vec R$ coordinates. 

To solve exactly the resulting six-dimensional stationary Schr\"odinger equation $\hat{H}\Psi=E\Psi$ for the two-particle wave function $\Psi$ with the corresponding two-particle energy $E$, we perform numerical computations in terms of the relative and CoM coordinates by applying a two-step approach \cite{Supplemental_Material}. First, we compute a set of basis functions, consisting of the CoM wave functions, being the eigenfunctions of the shifted radial harmonic oscillator with Dirichlet boundary condition, and the relative wave functions, being the eigenfunctions of a radially symmetric harmonic oscillator, but obeying the boundary conditions given by Eqs. \eqref{eqn:bc_BP_r12_1} and \eqref{eqn:bc_BP_r12_2}. Second, we represent the Hamiltonian $\hat{H}$ in terms of these radial basis functions, combined with spherical harmonics for the angular degrees of freedom. The resulting matrix is finally diagonalized. The use of these basis functions is  
very efficient, as only a few of them are actually required to obtain the exact solution.

To interpret correctly the solutions featuring a strong coupling between CoM and relative motions, it is essential to find and work within the coordinates where the couplings between different degrees of freedom are minimal, allowing for an effective reduction of the dimensionality of the problem. For our physical system, this is achieved with the hyperspherical coordinates \cite{Das2015}: the hyperradius $\xi=\sqrt{\frac{1}{2}r^2+2R^2}$, the hyperangle $\chi=\arctan(r/2R)$, the spherical angles $\theta$ and $\phi$ of $\vec r$, as well as $\Theta$ and $\Phi$ of $\vec R$. In the limit of a thin shell, i.~e. $r_0/a_\text{ho}\to \infty$, $\chi$ converges to $\theta_{12}/2$, with $\theta_{12}$ being the angle between the particle coordinate vectors $\vec{r}_1$ and $\vec{r}_2$. The explicit expressions of Eqs.~(1) and (2) in the hyperspherical coordinates, as well as the details of the subsequently employed approximations are presented in \hyperref[sec:endmatter]{End Matter}. Working with these hyperspherical coordinates, we can approximately separate the angular variables, $\theta$, $\phi$, $\Theta$ and $\Phi$, from $\xi$ and $\chi$. Furthermore, by considering the hyperradius $\xi$ as an adiabatic coordinate \cite{Supplemental_Material}, we can obtain approximately the hyperangular eigenfunctions $V_{n_\chi L\ell,\xi}(\chi)$ and the corresponding eigenvalues $\lambda(n_\chi, L,\ell;\xi)$. Here $L$ and $\ell$ are the angular momentum of CoM and relative motions, respectively, whereas $n_\chi$ denote the number of nodes of the wave function $V_{n_\chi L\ell,\xi}(\chi)$ along the hyperangular directions. These results allow us to find then the hyperradial wave functions $U_{n_\xi n_\chi L\ell}(\xi)$ and the two-particle energies $E_{n_\xi n_\chi L\ell}$, where $n_\xi$ represents the number of nodes of the wave function $U_{n_\xi n_\chi L\ell}(\xi)$ along the hyperradial direction. 
 
\begin{figure}
    \centering
\includegraphics{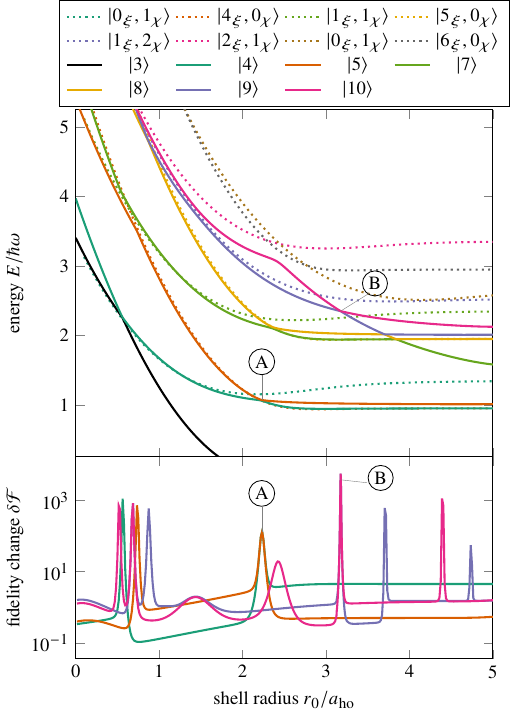}
    \caption{(color online) The exact ($E_n$) and approximated ($E_{n_\xi n_\chi}$) two-particle energies (upper panel) and fidelity change $\delta{\mathcal F}_n$  (lower panel) as functions of the shell radius $r_0$ for selected low energy states with zero total angular momentum ($J=0$ and $M_J=0$) and the scattering length $a_0=0.53a_\text{ho}$. We show the energies obtained from the exact six-dimensional computation (solid lines) and approximate approach based on the hyperspherical coordinates (dotted lines). The two marked avoided crossings 
    $\circled{A}$: $\ket{0_ \xi1_\chi}$ (dark green) with $\ket{4_\xi0_\chi}$ (orange), and $\circled{B}$: $\ket{1_\xi2_\chi}$ (purple) with  $\ket{0_\xi4_\chi}$ (pink), are analyzed in detail. The low-energy spectrum  including the case $J>0$ is presented in \cite[Fig. S.1]{Supplemental_Material}, where we also show that AC \circled{A} is the energetically lowest one for the given $a_0$.}
    \label{fig:fidelity_a0.5}
\end{figure}

\paragraph{Energy Spectrum.--} We begin our analysis by noting that for $r_0>0$ both angular momentum operators of the center-of-mass $\hat{\boldsymbol L}$ and the relative $\hat{\boldsymbol{\ell}}$ motions do not commute with the Hamiltonian of the system including the Fermi-Huang pseudopotential, or the corresponding boundary conditions Eqs. \eqref{eqn:bc_BP_r12_1} and \eqref{eqn:bc_BP_r12_2}. However, the operator of the total angular momentum $\hat{{\boldsymbol J}}=\hat{\boldsymbol L}+\hat{\boldsymbol{\ell}}$ does commute and therefore provides the useful quantum numbers $\{J, M_J\}$.

We first consider the case of the two-particle states with zero total angular momentum, $J=0$ and $M_J=0$. For the scattering length $a_0=0.53a_\text{ho}$, we present in the upper panel of \cref{fig:fidelity_a0.5} by solid lines the dependence of the exactly obtained two-particle energies $E_n$ on the shell radius $r_0$ \cite{Supplemental_Material}. We note that at $r_0=0$ the displayed energies $E_n(0)$ correspond to the two-particle states with $L=0$ and $\ell=0$. This follows from the fact that at $r_0=0$ there is no coupling between the center-of-mass and relative motion, and hence the operators $\hat{{\boldsymbol L}}$ and $\hat{{\boldsymbol \ell}}$ define the quantum numbers $L$ and $\ell$. For comparison we also display in the upper panel of \cref{fig:fidelity_a0.5} by dotted lines the energies $E_{n_\xi n_\chi 00}(r_0)\equiv E_{n_\xi n_\chi}(r_0)$ of the approximate states $\ket{n_\xi n_\chi 00}$, where the coupling of the angular momenta $\hat{{\boldsymbol L}}$ and $\hat{{\boldsymbol \ell}}$ is neglected, as described in \hyperref[sec:endmatter]{End Matter}. The neglect of the coupling between $\hat{{\boldsymbol L}}$ and $\hat{{\boldsymbol \ell}}$, determined by $\mathcal V_\text{c}$, Eq. \eqref{eqn:Vc} is the dominant reason for the deviation of $E_{n_\xi n_\chi}(r_0)$ from $E_n(r_0)$.

Starting from $r_0=0$ the energies $E_{n_\xi n_\chi}(r_0)$ and $E_{n}(r_0)$ decrease rapidly with an increasing shell radius $r_0$ and then saturate. This can be related to the transition of a three-dimensional harmonic oscillator to an effectively one-dimensional one for sufficiently large shell radii. Indeed, at $r_0=0$ the two-body spectrum is exactly given by the spectrum of a three-dimensional harmonic oscillator
\begin{equation}
    E_{n_\xi n_\chi}=\left[\mathcal{E}_{n_\chi}(a_0)+ 2n_\xi  +  \frac{3}{2}\right]\hbar \omega
    \label{eqn:bound_state-r0=0}
\end{equation}
shifted by the dimensionless energies $\mathcal{E}_{n_\chi}$ of the hyperangular motion, which is reduced to the relative motion for $r_0=0$. The hyperangular energies are determined by the roots of the transcendental equation \cite{Busch1998}
\begin{equation}
 \sqrt{2}\frac{\Gamma \left(\frac{3}{4}-\frac{1}{2}{\mathcal E_{n_\chi}}\right)}{\Gamma \left(\frac{1}{4} -\frac{1}{2}{\mathcal E_{n_\chi}}\right)} = -\frac{a_\text{ho}}{a_0}.
    \label{eqn:relE_alpha_dep}
\end{equation}
For $0<a_0/a_\text{ho}<1$ and $n_\chi=0$, we have obtained \cite{Supplemental_Material} $\mathcal E_{0} = -(a_\text{ho}/a_0)^2 + (1/8)(a_0/a_\text{ho})^2 + O\left[\left(a_0/a_\text{ho}\right)^6\right]$.

However, for $r_0\gg a_\text{ho}$, the bound-state energies feature the spectrum of the one-dimensional harmonic oscillator \cite{Supplemental_Material}
\begin{equation}
    E_{n_\xi 0}\approx\left[\mathcal{E}_{0}(a_0) + n_\xi +\frac12 \right]\hbar\omega, 
    \label{eqn:bound_state-r0=inf}
\end{equation}
that is also shifted by the hyperangular bound-state energy $\mathcal E_{0} = -(a_\text{ho}/a_0)^2 + (1/24)(a_0/a_\text{ho})^2 + O\left[\left(a_0/a_\text{ho}\right)^6\right]$, valid for $0<a_0/a_\text{ho}<1$.

Moreover, for intermediate shell radii $r_0$ the energy spectrum displays many crossings and avoided crossings (AC). In general, their positions and widths are determined by both $r_0$ and $a_0$. To identify and quantify ACs, corresponding to CIRs \cite{CIR,PhysRevA.94.022713}, we analyze the fidelity change \cite{Plotz2011,PloetzThesis} 
\begin{equation} 
    \delta {\mathcal F}_n(r_0)=\frac{1-\braket{n(r_0)}{n(r_0+\delta r_0)}}{(\delta r_0/a_\text{ho})^2}\label{eqn:fidelity_change}
\end{equation}
of the exact eigenstates $|n\rangle$ as a function of $r_0$ for a given $a_0$. As clearly shown in the lower panel of \cref{fig:fidelity_a0.5}, $\delta{\mathcal F}_n(r_0)$ exhibits peaks.
When the peaks of the fidelity change $\delta{\mathcal F}_{n_1}$ of the state $\ket{n_1}$ and $\delta{\mathcal F}_{n_2}$ of the state $\ket{n_2}$ coincide, the constituent states of the AC are identified. Further, we can determine the position and width of the AC from these peaks.

\begin{figure}
    \centering
\includegraphics{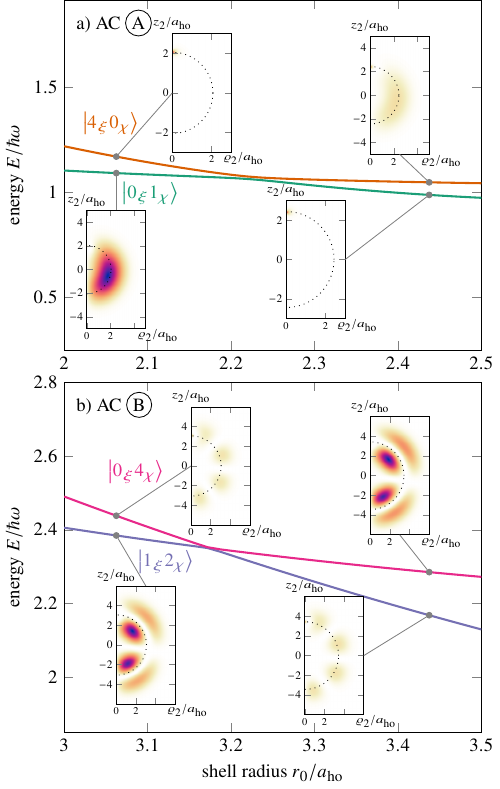}
    \caption{(color online) Magnifications of the energy levels at the labeled avoided crossings $\circled{A}$ and $\circled{B}$, \cref{fig:fidelity_a0.5}.  As insets, the conditional one-particle probability density $P_{2|1}(\varrho_2,z_2)$, \cref{eqn:P21}, of the involved states before and after the avoided crossing are shown. In panel a), the avoided crossing $\circled{A}$ couples the molecule state $\ket{0_\xi1_\chi}$ with the excited trap state $\ket{4_\xi0_\chi}$ and represents a bound-trap CIR.  In panel b) the trap state $\ket{0_\xi4_\chi}$ with excitation in $\chi$ direction features  the avoided crossing $\circled{B}$ with the trap state $\ket{1_\xi2_\chi}$ having excitations in both $\xi$ and $\chi$ directions, visualized here by the nodes in the radial and polar direction. The dotted half circle represents in all insets the minimum of the trapping potential. See also Fig.~S2 and Fig.~S3 in \cite{Supplemental_Material} for density plots of the states as functions of $r$ and $R$.}
    \label{fig:ac_a=0.5}
\end{figure}

\paragraph{Structure of CIRs.--}
We now consider in detail two kinds of avoided crossings, labeled by $\circled{A}$ and $\circled{B}$ in \cref{fig:fidelity_a0.5}, respectively.
To visualize the states involved at the ACs, we introduce the conditional probability density for the second particle of state $\ket{n}$
\begin{equation}
    P_{2|1}^{(n)}(\varrho_2,z_2)=2\pi\rho\abs{\Psi_n(\rho_1=0,z_1=r_0,\varrho_2, z_2)}^2, \label{eqn:P21}
\end{equation}
given that the first particle is located at the potential minimum on the north pole. Exploiting the cylindrical symmetry of the selected states we have introduced the cylindrical coordinates $(\varrho_i, z_i)$ for the $i$-th particle, suppressing the dispensable azimuthal angles $\phi_i$. 

In~\cref{fig:ac_a=0.5}~a), we display AC $\circled{A}$, which couples the diabatic states $\ket{4_\xi0_\chi}$, shown in the top left inset by the density plot of $P_{2|1}^{(5)}(\varrho_2,z_2)$,
and $\ket{0_\xi1_\chi}$, shown in the bottom left inset by the density plot of $P_{2|1}^{(4)}(\varrho_2,z_2)$.
The state $\ket{4_\xi0_\chi}$ is tightly localized at the north pole, which can be attributed to a small relative distances $r$ to the first particle. Actually, it resembles a bound state of two non-trapped particles with the characteristic exponential decay $\exp(-r/a_0)$. Such states correspond to \emph{molecules on the shell} and only appear for positive $a_0$. They exhibit the molecular form only for small positive values of $a_0/a_\text{ho}$, since otherwise the trap potential dominates.
The second state $\ket{0_\xi1_\chi}$ of AC $\circled{A}$, possesses an approximate hyperradial symmetry, which is dictated by the trap structure and depicted in \cite[Fig.~S2]{Supplemental_Material}. The density plot in \cref{fig:ac_a=0.5}~a) also shows the approximate radial symmetry, while the hyperangular node appears as a node in the polar direction on the sphere.  We denote such states as \emph{trap states} of two particles. Avoided crossings, such as $\circled{A}$, describe an interplay between the molecular and trap states, are named here as \emph{bound-trap} CIRs.
\begin{figure}
    \centering
\includegraphics{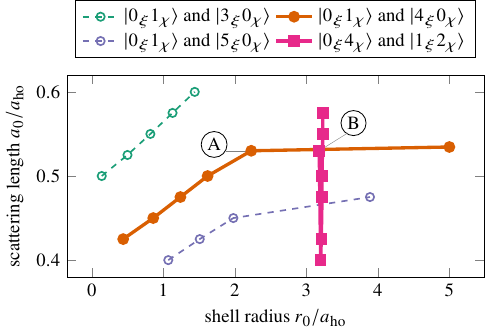}
    \caption{Positions of selected avoided crossings of low-energy states with zero total angular momentum ($J=0$ and $M_J=0$) in the $a_0$-$r_0$ parameter space. The circles mark bound-trap ACs, similar to $\circled{A}$ (orange). Due to strong dependence of energy of the molecular state on $a_0$, a given bound-trap AC only appears for a narrow interval of $a_0$ before another AC with the next lower (turquoise), or higher (purple) trap state arises, taking finally its place. In contrast, trap-trap ACs (squares), similar to $\circled{B}$, dependent weakly on $a_0$ and strongly on $r_0$, resulting from the weak dependence on $a_0$ and strong dependence $r_0$ of the energy levels spacing of the involved states.}
    \label{fig:ac_position} 
\end{figure}

The second kind of CIRs is exemplified by AC $\circled{B}$, which is displayed in \cref{fig:ac_a=0.5} b). Here, the diabatic state $\ket{0_\xi4_\chi}$, depicted in the top left inset by the density plot of 
$P_{2|1}^{(10)}(\varrho_2,z_2)$,
is coupled to the state $\ket{1_\xi2_\chi}$, depicted in the bottom left inset by the density plot of 
$P_{2|1}^{(9)}(\varrho_2,z_2)$. The number of hyperangular nodes is also translated into the polar direction. Moreover, the nodes of the hyperradial dimension appear as nodes in the spherical radius $r_2$.  Since neither of the involved states is a  molecular one, we call such ACs  \emph{trap-trap} CIRs. We emphasize that both kinds of resonances originate entirely from the strong coupling between the relative and CoM degrees of freedom.

In \cref{fig:ac_position} the positions of the AC $\circled{A}$ and other two bound-trap CIRs (circles), as well as the AC $\circled{B}$ (squares) are presented in the parameter space of the scattering length $a_0$ and the shell radius $r_0$. The positions of the bound-trap CIRs strongly depend on $a_0$, as the energy of the molecular state and hence the level spacing of the involved states are crucially determined by the value of $a_0$. Consequently, a given trap state avoids crossing with the corresponding molecular state only in a small interval of $a_0$ and with increasing $a_0$ the position of this AC becomes shifted towards larger shell radii until it disappears. Note that close to the vanishing of an AC the curve in the $r_0$-$a_0$ plane becomes nearly horizontal (here only well resolved for the orange AC). This behavior is due to the fact that the involved energy levels become nearly parallel for large $r_0$, as depicted in \cref{fig:fidelity_a0.5}.
In contrast, the positions of trap-trap ACs like AC $\circled{B}$ are nearly independent of $a_0$, as the shape and the energy level spacing of the involved trap states are essentially unaffected by a change of the scattering length. 

We emphasize that the avoided crossings are not restricted to the case $J=0$ and $M_J=0$. Indeed, they also occur in the two-article energy spectrum for $J>0$, see \cite[Fig. S1]{Supplemental_Material}.

\paragraph{Summary and outlook.--}

We have computed the energy spectrum and corresponding wave functions of two bosonic particles, which are confined in a spherically symmetric shell-shaped trap and interact with each other through the three-dimensional $s$-wave zero-range potential. Further, we have identified two types of CIRs as avoided crossings between (i) the bound (molecular) state with strong localization in the relative coordinate and CoM excitation and the trap (non-molecular) state without CoM excitation, and between (ii) two trap states. We have quantitatively determined  the positions and widths of these resonances and shown clearly that they are induced by the strong coupling of the relative and CoM motions of the two particles in the shell-shaped trap. 

The found CIRs could be probed experimentally by observing losses in atomic bubbles. In fact, an adiabatic change of the shell radius $r_0$ around a bound-trap CIR drives the formation of molecules in the shell-shaped trap. The subsequent collisions of the formed molecules with each other, or with other atoms lead to the formation of tighter molecules and atoms with high kinetic energy, resulting in the loss of atoms from the trap. 
Depending on the actual realization of the quantum bubble, there are different schemes to scan the shell radius in a controllable way. In the case of the rf-dressing technique \cite{ZobayGarraway2001,*ZobayGarraway2004,Garraway_2016,PERRIN2017181}, this can easily be done by scanning the rf-detuning $\Delta$, due to the fact that $r_0\propto \sqrt{\Delta}$. In the case of a shell potential generated with a two-species atomic mixture \cite{PhysRevA.106.013309,Jia.2022}, the time-dependent shell radius can be realized, for example, during free expansion.  
In addition, it would be of great interest to explore CIRs that occur in (i) ellipsoid-shaped traps, resulting usually from applying the rf-dressing scheme \cite{ZobayGarraway2001,*ZobayGarraway2004,Garraway_2016,PERRIN2017181}, and (ii) ring-shaped quantum gases \cite{RevModPhys.94.041001}. The latter has useful applications in building compact and stable sensors for inertial and non-inertial forces.   

We thank A. Wolf for fruitful discussions and helpful suggestions. The authors acknowledge support by the state of Baden-Württemberg through bwHPC and the German Research Foundation (DFG) through grant no INST 40/575-1 FUGG (JUSTUS 2 cluster).

\bibliography{references}

\appendix
\section{End Matter}\label{sec:endmatter}
In hyperspherical coordinates the Hamiltonian, \cref{eqn:H}, reads
\begin{equation}
    \hat {H}=-\frac{\hbar^2}{2m}\left(\pdv[2]{\xi} +\frac{5}{\xi}\pdv{\xi} -\frac{\hat\Lambda^2}{\xi^2}\right) + \frac{m\omega^2}{2}(\xi-\xi_0)^2+\mathcal V_\text{c}, \label{eqn:H_hs}
\end{equation}
where $\xi_0=\sqrt{2}r_0$ and the operator $\hat\Lambda^2=-\hat\Lambda^2_0 +\mathcal W_\xi(\chi)$ consists of the Laplacian
\begin{equation}
    \hat \Lambda_0^2 = \pdv[2]{\chi} + 4\cot(2\chi)\pdv{\chi} -\frac{\hat\ell^2}{\sin[2](\chi)} -\frac{\hat L^2}{\cos[2](\chi)}
\end{equation}
on the six-dimensional hypersphere with the relative $\hat \ell^2$ and CoM $\hat L^2$ orbital angular momentum operators, and the potential
\begin{equation}
    \mathcal W_\xi(\chi)= \frac{2\xi^3\xi_0}{a_\text{ho}^4}\left[1-\frac{2+\sin[2](2\abs{\frac \pi 4-\chi})}{3\cos(\frac \pi 4-\abs{\frac \pi 4-\chi})} \right]
\end{equation}
with the characteristic length $a_\text{ho}\equiv \sqrt{\hbar/(m\omega)}$ of the shell-shaped trapping potential $V_0({\bf r})$, defining also the shell width.  

The last term in Eq. (\ref{eqn:H_hs})
\begin{gather}
     \mathcal V_\text{c}= -m\omega^2\xi \xi_0 \sum_{l=1}^\infty\frac{\sin[2l](\frac{\pi}{4}-\abs{\chi-\frac\pi4})}{\cos[2l+1](\frac{\pi}{4}-\abs{\chi-\frac\pi4})}\left[
    \frac{\sin[2](\frac{\pi}{4}-\abs{\chi-\frac\pi4})}{4l+3} \right. \notag\\
    \left.-\frac{\cos[2](\frac{\pi}{4}-\abs{\chi-\frac\pi4})}{4l-1}
    \right]
\frac{4\pi}{4l+1} \sum_{\mu=-2l}^{2l}
     Y_{2l,\mu}(\theta,\phi)Y_{2l,\mu}^*(\Theta,\Phi),
     \label{eqn:Vc}
\end{gather}
with $Y_{l,m}$ being the spherical harmonics on the three-dimensional sphere, couples all degrees of freedom.

Further, the boundary conditions, Eqs. \eqref{eqn:bc_BP_r12_1} and \eqref{eqn:bc_BP_r12_2}, take a simpler form in hyperspherical coordinates:
\begin{subequations}
 \begin{align}
    \frac{1}{\sin(2\chi)\Psi}\frac{\mathrm d}{\mathrm d\chi} [\sin(2\chi)\Psi]\Big|_{\chi\to0}&=-\frac{\sqrt 2\xi}{a_0}\;\;\;{\rm for}\;\ell=0\label{eqn:bc_BP_chi_l0}\\  
     \sin(2\chi)\Psi\Big|_{\chi\to0}&=0 \;\;\;{\rm for}\;\ell>0.  \label{eqn:bc_BP_chi_l>0}
 \end{align}
 \end{subequations}

When we neglect the coupling term $\mathcal V_\text{c}$, the angular variables of the relative ($\theta$ and $\phi$) and CoM ($\Theta$ and $\Phi$) motions can be separated from each other as well as from the hyperradius $\xi$ and hyperangle $\chi$ both in the Hamiltonian, \cref{eqn:H_hs}, and the boundary conditions, \cref{eqn:bc_BP_chi_l0,eqn:bc_BP_chi_l>0}. 
In this way, the Hamiltonian \cref{eqn:H_hs} would have the form of a spherically symmetric six-dimensional radially shifted harmonic oscillator, if the operator $\hat \Lambda^2$ and the boundary condition,
\cref{eqn:bc_BP_chi_l0},
were $\xi$-independent. However, this is not the case and we have to use a proper scheme to solve the reduced two-dimensional Schr\"odinger equation numerically. This is done approximately \cite{Supplemental_Material} within the adiabatic approximation. Namely, we first compute the eigenvalues $\lambda(n_\chi, L,\ell;\xi)$ and eigenfunctions $V_{n_\chi L\ell,\xi}(\chi)$ of the operator $\hat\Lambda^2$, taking into account \cref{eqn:bc_BP_chi_l0,eqn:bc_BP_chi_l>0},
with $\xi$ considered as a parameter. We then replace $\hat\Lambda^2$ in \cref{eqn:H_hs} by $\lambda$, neglect the coupling $\mathcal V_c$ \footnote{The neglect of $\mathcal V_c$ is valid only for $r_0 \lesssim 2a_\text{ho}$. For increasing values of $r_0$, more and more terms of $\mathcal V_c$ have to be taken into account. For $r_0/a_\text{ho} \to \infty$, however, the sum of $\mathcal W/\xi^2+\mathcal V_c$ vanishes for particles exactly on the sphere with radius $r_0$.}, and solve numerically the obtained eigenvalue equation for the hyperradial wave functions $U_{n_\xi n_\chi L\ell}(\xi) $ and the corresponding two-particle energies $E_{n_\xi n_\chi L\ell}$.

\clearpage



\section*{Supplementary material}

# Supplemental Material to: Confinement Induced Resonances in Spherical Shell Traps

C. Moritz Carmesin[1, *] and Maxim A. Efremov[1, 2]

[1]*Institut für Quantenphysik and Center for Integrated Quantum Science and Technology (IQST), Universität Ulm, 89081 Ulm, Germany*
[2]*German Aerospace Center (DLR), Institute of Quantum Technologies, 89081 Ulm, Germany*

## NUMERICAL SOLUTION

In order to obtain the exact numerical solution of the stationary Schrödinger equation for the two-particle wave function, we use the spherical coordinate representation for the relative $r$ and CoM $R$ coordinates scaled in terms of the oscillator length $a_{\mathrm{ho}} = \sqrt{\hbar/(m\omega)}$, that is $r = \{a_{\mathrm{ho}}\rho, \theta, \phi\}$ and $R = \{a_{\mathrm{ho}}\mathcal{R}, \Theta, \Phi\}$. The Hamiltonian then reads

$$\hat{H} = -\frac{1}{4}\nabla_{\mathcal{R}}^2 - \nabla_{\rho}^2 + \mathcal{R}^2 + \frac{1}{4}\rho^2 - s\left(\sqrt{\mathcal{R}^2 + \frac{1}{4}\rho^2 + \mathcal{R}\rho\cos\theta_{r\mathcal{R}}} + \sqrt{\mathcal{R}^2 + \frac{1}{4}\rho^2 - \mathcal{R}\rho\cos\theta_{r\mathcal{R}}}\right) + s^2, \tag{S.1}$$

where $s = r_0/a_{\mathrm{ho}}$ and $\theta_{r\mathcal{R}}$ is the angle between $r$ and $R$. By representing the sum of the two square roots [1]

$$w \equiv \sqrt{\mathcal{R}^2 + \frac{1}{4}\rho^2 + \mathcal{R}\rho\cos\theta_{r\mathcal{R}}} + \sqrt{\mathcal{R}^2 + \frac{1}{4}\rho^2 - \mathcal{R}\rho\cos\theta_{r\mathcal{R}}} = \sum_{l=0}^{\infty}\left(\frac{2a^{2l}}{b^{2l+1}}\right)\left(\frac{a^2}{4l+3} - \frac{b^2}{4l-1}\right)P_{2l}(\cos\theta_{rR}) \tag{S.2}$$

in terms of the Legendre polynomials $P_{2l}(\cos\theta_{r\mathcal{R}})$, with $a = \min\{\frac{1}{2}\rho, \mathcal{R}\}$ and $b = \max\{\frac{1}{2}\rho, \mathcal{R}\}$, we can recast the Hamiltonian into the form

$$\hat{H} = \hat{H}_0 + s\hat{W} = \hat{H}_{\mathcal{R}} + \hat{H}_{\rho} + s\hat{W} \tag{S.3}$$

with

$$\hat{H}_{\mathcal{R}} = -\frac{1}{4}\left(\frac{\partial^2}{\partial\mathcal{R}^2} + \frac{2}{\mathcal{R}}\frac{\partial}{\partial\mathcal{R}} - \frac{\hat{L}^2}{\mathcal{R}^2}\right) + (\mathcal{R}-s)^2 \tag{S.4}$$

$$\hat{H}_{\rho} = -\left(\frac{\partial^2}{\partial\rho^2} + \frac{2}{\rho}\frac{\partial}{\partial\rho} - \frac{\hat{\ell}^2}{\rho^2}\right) + \frac{1}{4}\rho^2 \tag{S.5}$$

$$\hat{W} = -w + 2\mathcal{R}, \tag{S.6}$$

where $\hat{L}$ and $\hat{\ell}$ are the angular momentum operators for the CoM and relative motion, respectively. In addition, the particle-particle $s$-wave interaction is assumed to be well described by the Bethe-Peierls boundary condition

$$\frac{1}{\rho\Psi(\rho,\mathcal{R})}\frac{\mathrm{d}}{\mathrm{d}\rho}[\rho\Psi(\rho,\mathcal{R})]\bigg|_{\rho\to 0} = -\frac{a_{\mathrm{ho}}}{a_0} \quad \text{for} \quad \ell = 0 \quad \text{and} \quad \rho\Psi(\rho,\mathcal{R})|_{\rho\to 0} = 0 \quad \text{for} \quad \ell > 0 \tag{S.7}$$

for the two-particle wave function $\Psi(\rho, \mathcal{R})$, where $a_0$ is the $s$-wave scattering length of the particle-particle interaction. Next, we determine and use the eigenfunctions of $\hat{H}_{\mathcal{R}}$ and $\hat{H}_{\rho}$ as basis functions to finally compute the eigenenergies and eigenstates of the two-particle Hamiltonian $\hat{H} = \hat{H}_0 + s\hat{W}$.

### CoM Basis Functions

The angular part of the CoM Hamiltonian $\hat{H}_{\mathcal{R}}$ can be separated by recalling that the eigenfunctions of the angular momentum operator $\hat{L}^2$ are the spherical harmonics $Y_{LM}(\Theta, \Phi)$ with the eigenvalues $L(L+1)$ and $L = 0, 1, 2, \ldots$. Hence, the full eigenfunctions of $\hat{H}_{\mathcal{R}}$ are given by $\psi_{\mathrm{CoM}}(\mathcal{R}, \Theta, \Phi) = (1/\mathcal{R})u_{n_{\mathcal{R}}L}(\mathcal{R})Y_{LM}(\Theta, \Phi)$, where the functions $u_{n_{\mathcal{R}}L}$ obey the radial Schrödinger equation

$$-\frac{1}{4}\frac{\mathrm{d}^2 u_{n_{\mathcal{R}}L}}{\mathrm{d}\mathcal{R}^2} + \left[\frac{L(L+1)}{4\mathcal{R}^2} + (\mathcal{R}-s_0)^2\right]u_{n_{\mathcal{R}}L}(\mathcal{R}) = \mathcal{E}_{n_{\mathcal{R}}L}u_{n_{\mathcal{R}}L}(\mathcal{R}). \tag{S.8}$$

In general, $s_0$ may be set equal to $s$, but for our numerical computations, it is advantageous, to use also different values, see below. The solutions of Eq. (S.8) can be given in terms of the biconfluent Heun functions [2, 3]. However, we do not make use of this fact

and instead directly solve the eigenvalue problem using the ApproxFun Julia package [4] leveraging a Chebyshev-polynomial-based spectral method.

In the case of $s_0 = 0$ we reproduce the case of the spherically symmetric three-dimensional harmonic oscillator with the eigenenergies

$$\mathcal{E}_{n_\mathcal{R} L} = \frac{3}{2} + 2n_\mathcal{R} + L \tag{S.9}$$

and the corresponding wave function

$$u_{n_\mathcal{R} L}(\mathcal{R}) = A_{n_\mathcal{R} L} \mathcal{R}^{L+1} e^{-\mathcal{R}^2} L_{n_\mathcal{R}}^{(L+\frac{1}{2})}\left(2\mathcal{R}^2\right) \tag{S.10}$$

represented in terms of the Laguerre polynomials $L_n^{(\alpha)}(z)$, with $A_{n_\mathcal{R} l}$ being the normalization constant.

### Relative Basis Functions

Analogously to the CoM case, the eigenfunctions of $\hat{H}_\rho$ are given by $\varphi(\rho, \theta, \phi) = (1/\rho) v_{n_\rho \ell}(\rho) Y_{\ell m}(\theta, \phi)$, where the functions $v_{n_\rho \ell}(\rho)$ obey the radial Schrödinger equation

$$-\frac{d^2 v_{n_\rho \ell}(\rho)}{d\rho^2} + \left[\frac{1}{4}\rho^2 + \frac{\ell(\ell+1)}{\rho^2}\right] v_{n_\rho \ell}(\rho) = \mathcal{E}_{n_\rho \ell} v_{n_\rho \ell}(\rho) \tag{S.11}$$

and are subject to the boundary conditions

$$\left.\frac{1}{v_{n_\rho \ell}(\rho)} \frac{d}{d\rho} v_{n_\rho \ell}(\rho)\right|_{\rho \to 0} = -\frac{a_{\text{ho}}}{a_0} \quad \text{for} \quad \ell = 0 \tag{S.12a}$$

and

$$v_{n_\rho \ell}(0) = 0 \quad \text{for} \quad \ell > 0. \tag{S.12b}$$

For $\ell > 0$ we again reproduce the spherically symmetric three-dimensional harmonic oscillator with the eigenenergies

$$\mathcal{E}_{n_\rho \ell}(\rho) = \left(\frac{3}{2} + 2n_\rho + \ell\right) \quad (\ell > 0) \tag{S.13}$$

and the wave functions

$$v_{n_\rho \ell}(\rho) = A_{n_\rho \ell} \rho^{\ell+1} e^{-\frac{\rho^2}{4}} L_{n_\rho}^{(\ell+\frac{1}{2})}\left(\frac{\rho^2}{2}\right) \quad (\ell > 0), \tag{S.14}$$

where $A_{n_\rho \ell}$ is the normalization constant.

In the case $\ell = 0$, the solutions are best written in terms of the parabolic cylinder functions $D_\nu(x)$

$$v_{n_\rho 0}(\rho) = A_{n_\rho 0}(a_0) D_{\mathcal{E}_{n_\rho 0}(a_0) - \frac{1}{2}}(\rho), \tag{S.15}$$

where $A_{n_\rho 0}(a_0)$ is the normalization constant, whereas the energies $\mathcal{E}_{n_\rho 0}(a_0)$ are determined as the roots of the transcendental equation

$$\sqrt{2} \frac{\Gamma\left(\frac{3}{4} - \frac{1}{2}\mathcal{E}_{n_\rho 0}\right)}{\Gamma\left(\frac{1}{4} - \frac{1}{2}\mathcal{E}_{n_\rho 0}\right)} = -\frac{a_{\text{ho}}}{a_0}. \tag{S.16}$$

This equation can be derived by inserting the asymptotic expansion of $v_{n_\rho 0}(\rho)$ as $\rho \to 0$, Eq. (S.15), into Eq. (S.12a), or by the method used in the original paper [5]. Due to numerical difficulties at the evaluation of the parabolic cylinder functions for negative orders $\nu < 0$, we resort to the numerical solution of the eigenvalue problem, Eqs. (S.11) and (S.12), with the help of the Julia ApproxFun package [4]. For $0 < a_0/a_{\text{ho}} < 1$ we obtain from Eq. (S.16) the asymptotic expansion for the ground state energy

$$\mathcal{E}_{00} = -\frac{1}{\alpha_0^2} + \frac{\alpha_0^2}{8} + \frac{7\alpha_0^6}{128} + O\left(\alpha_0^{10}\right). \tag{S.17}$$

Here made use of the asymptotic expansion of the Gamma function for large arguments and the interative method to solve the equation.

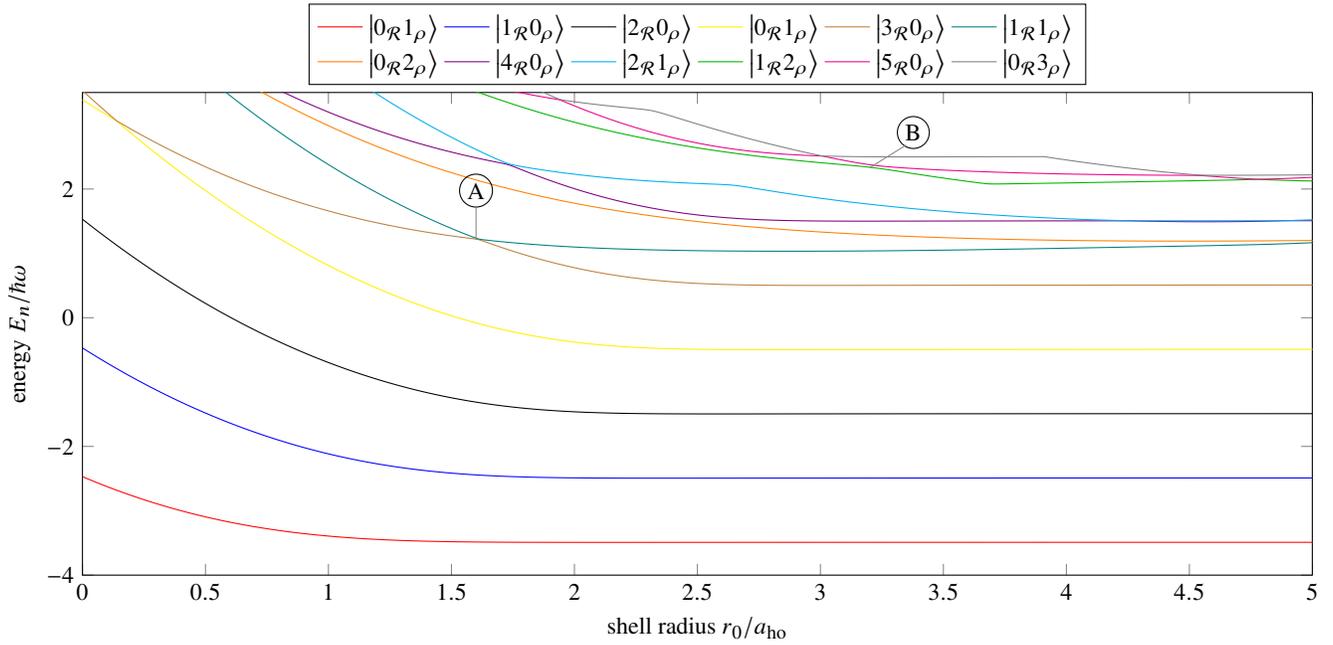

Figure S.1. Spectrum of the two-particle Hamiltonian Eq. (S.1) for zero total angular momentum, that is with $L = \ell = 0$, as a function of the shell radius $r_0$ obtained by the exact numerical diagonalization. For the labeling of the energies, we use the quantum numbers $n_{\mathcal{R}}$ and $n_\rho$ of the corresponding states $|n_{\mathcal{R}}, n_\rho\rangle$ at $r_0 = 0$. The states with the lowest energies contain the ground state of the relative motion, corresponding to a molecule on the shell, and show the spectrum of an harmonic oscillator with constant distance between the energy levels for large shell radii $r_0 \gg a_{\text{ho}}$. Two symbols Ⓐ and Ⓑ identify the avoided crossings discussed in the Letter.

### Diagonalizaion

Having obtained the functions $u_{n_{\mathcal{R}} L}(\mathcal{R})$ and $v_{n_\rho \ell}(\rho)$, we are in the position to expand the coupling operator $\hat{W}$, Eq. (S.6), in $\hat{H}$ in terms of these functions. We make use of the fact that each summand of $\hat{W}$ factors in a radial and angular part, cf. Eq. (S.2). The matrix elements of the latter can be given exactly in terms of the Clebsch-Gordon coefficients, while the matrix elements determined by the functions of $\mathcal{R}$ and $\rho$ are obtained by two-dimensional integration based on the Gauss-Lobatto scheme.

In order to save computational time during scanning over different values of $s$, we do not compute new basis functions and even more expensive matrix elements for each value of $s$. Instead, we just compute them for fixed values $s_0$ and rewrite the original Hamiltonian in the form

$$\hat{H} = \hat{H}_0 + s\hat{W} - 2\hat{\mathcal{R}}\delta s + 2s_0 \delta s + (\delta s)^2 \tag{S.18}$$

with $\delta s = s - s_0$. Now have to additionally calculate the matrix elements of the CoM position operator $\hat{\mathcal{R}}$. In the chosen basis only $\hat{W}$ and $\hat{\mathcal{R}}$ contain off-diagonal terms. We use the ARPACK library to calculate the eigenenergies and eigenfunctions of Eq. (S.18) in a matrix free fashion, by exploiting the sum-of-products structure of $\hat{W}$ and the tensor product structure of $\hat{\mathcal{R}}$.

In Fig. S.1, we show the spectrum of Eq. (S.1) for states with zero total angular momentum obtained by the diagonalization procedure described above. The states originating from the ground state of the relative motion, i. e. the bound state, show the spectrum of an one-dimensional harmonic oscillator with a constant level spacing of $\hbar\omega$, for sufficiently large $r_0$. However, for $r_0 = 0$ the energy spacing is $2\hbar\omega$, which corresponds to the spherical three-dimensional harmonic oscillator.

### APPROXIMATE SOLUTION

In order to obtain the approximate solutions for the eigenenergies and eigenfunctions, we use hyperspherical coordinates, namely the hyperradius $\xi = \sqrt{\frac{1}{2}r^2 + 2R^2}$, the hyperangle $\chi = \arctan(r/2R)$, the spherical angles $\theta$ and $\phi$ of $\mathbf{r}$, as well as $\Theta$ and $\Phi$ of $\mathbf{R}$. In these coordinates [6], the two-particle Hamiltonian reads

$$\hat{H} = -\frac{\hbar^2}{2m}\left(\frac{\partial^2}{\partial \xi^2} + \frac{5}{\xi}\frac{\partial}{\partial \xi} - \frac{\hat{\Lambda}^2}{\xi^2}\right) + \frac{m\omega^2}{2}(\xi - \xi_0)^2 + \mathcal{V}_c \tag{S.19}$$

with $\xi_0 = \sqrt{2}r_0$. Here, $\hat{\Lambda}^2 = -\hat{\Lambda}_0^2 + \mathcal{W}_\xi(\chi)$ consists of the Laplacian operator on the six-dimensional hypersphere

$$\hat{\Lambda}_0^2 = \frac{\partial^2}{\partial \chi^2} + 4\cot(2\chi)\frac{\partial}{\partial \chi} - \frac{\hat{\ell}^2}{\sin^2(\chi)} - \frac{\hat{L}^2}{\cos^2(\chi)} \tag{S.20}$$

and the potential

$$\mathcal{W}_\xi(\chi) = \frac{2\xi^3 \xi_0}{a_{ho}^4}\left[1 - \frac{2 + \sin^2(2|\frac{\pi}{4} - \chi|)}{3\cos(\frac{\pi}{4} - |\frac{\pi}{4} - \chi|)}\right] \tag{S.21}$$

depending on $\xi$ and $\chi$. The third term [1]

$$\mathcal{V}_c = -m\omega^2 \xi \xi_0 \sum_{l=1}^{\infty} \frac{\sin^{2l}(\frac{\pi}{4} - |\chi - \frac{\pi}{4}|)}{\cos^{2l+1}(\frac{\pi}{4} - |\chi - \frac{\pi}{4}|)}\left[\frac{\sin^2(\frac{\pi}{4} - |\chi - \frac{\pi}{4}|)}{4l+3} - \frac{\cos^2(\frac{\pi}{4} - |\chi - \frac{\pi}{4}|)}{4l-1}\right]\frac{4\pi}{4l+1}\sum_{\mu=-2l}^{2l} Y_{2l,\mu}(\theta,\phi)Y_{2l,\mu}^*(\Theta,\Phi) \tag{S.22}$$

in Eq. (S.19) connects all degrees of freedom.

By neglecting the coupling term $\mathcal{V}_c$ and using the spherical harmonics $Y_{LM}(\Theta, \Phi)$ and $Y_{\ell m}(\theta, \phi)$, we can separate the spherical angles $\theta$, $\phi$, $\Theta$, and $\Phi$ from the $\xi$ and $\chi$ variables and arrive at

$$\Psi_{n_\xi n_\chi LM\ell m}(\xi, \chi, \Theta, \Phi, \theta, \phi) = \frac{U_{n_\xi n_\chi L\ell}(\xi)}{\xi^{5/2}} \frac{V_{n_\chi L\ell,\xi}(\chi)}{\sin(2\chi)} Y_{LM}(\Theta, \Phi) Y_{\ell m}(\theta, \phi) \tag{S.23}$$

for the two-particle eigenfunction. Here the functions $V_{n_\chi L\ell,\xi}(\chi)$ obey the hyperangular Schrödinger equation

$$-\frac{\partial^2}{\partial \chi^2} V_{n_\chi L\ell,\xi}(\chi) + \left[\mathcal{W}_\xi(\chi) + \frac{L(L+1)}{\cos^2(\chi)} + \frac{\ell(\ell+1)}{\sin^2(\chi)} - 4\right] V_{n_\chi L\ell,\xi}(\chi) = \lambda V_{n_\chi L\ell,\xi}(\chi) \tag{S.24}$$

with the Bethe-Peierls boundary condition

$$\left.\frac{1}{V_{n_\chi L\ell,\xi}(\chi)}\frac{d}{d\chi}V_{n_\chi L\ell,\xi}(\chi)\right|_{\chi \to 0} = -\frac{\sqrt{2}\xi}{a_0} \quad \text{for} \quad \ell = 0. \tag{S.25a}$$

and

$$V_{n_\chi L\ell,\xi}(0) = 0 \quad \text{for} \quad \ell > 0. \tag{S.25b}$$

The hyperangular quantum number $n_\chi$ corresponds to the number of nodes in the hyperangular direction. In general, Eq. (S.24) with the conditions Eq. (S.25) can only be solved numerically. As with the basis functions for the exact diagonalization, we have used the Julia ApproxFun package [4] to obtain the eigenfunctions $V_{n_\chi L\ell,\xi}(\chi)$ with the corresponding eigenvalues $\lambda = \lambda(n_\chi, L, \ell, \xi)$.

Finally, we assume that the hyperangular wave function $V_{n_\chi L\ell,\xi}(\chi)$ changes only adiabatically with the hyperradius $\xi$, such that terms involving $\frac{\partial}{\partial \xi}V_{n_\chi L\ell,\xi}(\chi)$ can be neglected. This procedure is similar to the Born-Oppenheimer approximation [7] and allows us to calcuate the two-particle energy $E_{n_\xi n_\chi L\ell}$ from the corresponding hyperradial Schrödinger equation

$$-\frac{\hbar^2}{2m}\frac{\partial^2 U_{n_\xi n_\chi L\ell}(\xi)}{\partial \xi^2} + \left[\frac{\hbar^2}{2m}\frac{\lambda(n_\chi, L, \ell, \xi) + 15/4}{\xi^2} + \frac{m\omega^2}{2}(\xi - \xi_0)^2\right]U_{n_\xi n_\chi L\ell}(\xi) = E_{n_\xi n_\chi L\ell}U_{n_\xi n_\chi L\ell}(\xi) \tag{S.26}$$

for the function $U_{n_\xi n_\chi L\ell}(\xi)$.

To solve Eq. (S.26) numerically, we have used a sin-based Lagrange mesh [8, 3.7.2.] and then employed the ARPACK package. The mesh method allows us to restrict the number of required evaluations of $\lambda(n_\chi, \xi, L, \ell)$, in contrast to the previously applied spectral method.

In the limit of large shell radii, as well as for $L = \ell = 0$, we are able to derive approximate analytical expressions for the energies and the wave functions involving the hyperangular ground state. To do it, we expand the potential $\mathcal{W}_\xi(\chi)$, Eq. (S.21), around $\chi = 0$ and find

$$\mathcal{W}_\xi(\chi) = \frac{\xi^3 \xi_0}{3a_{ho}}\chi^2 + O(\chi^3). \tag{S.27}$$

Then Eq. (S.24) takes the form

$$\frac{\partial^2}{\partial \chi^2} V_0(\chi) + \left[ -\frac{\xi^3 \xi_0}{3 a_{\text{ho}}^4} \chi^2 + 4 + \lambda \right] V_0(\chi) = 0 \tag{S.28}$$

with the solutions

$$D_\nu \left[ \pm \left( \frac{2}{3} \xi^3 \xi_0 \right)^{\frac{1}{4}} \chi \right] \tag{S.29}$$

in terms of the parabolic cylinder functions, where the order $\nu$ is related to the eigenvalue $\lambda$

$$\lambda = -4 + \frac{(2\nu + 1)\sqrt{\xi^3 \xi_0}}{\sqrt{3} a_{\text{ho}}^2}. \tag{S.30}$$

The values of $\nu$ are then the roots of the transcendental equation

$$-\frac{\sqrt{2} \Gamma\left(\frac{1}{4}\left(3 - \sqrt{3}\nu\right)\right)}{\sqrt[4]{3} \Gamma\left(\frac{1}{4}\left(1 - \sqrt{3}\nu\right)\right)} = \frac{a_{\text{ho}}}{a_0}, \tag{S.31}$$

resulting from the Bethe-Peierls boundary condition Eq. (S.25a).

For $0 < \frac{a_0}{a_{\text{ho}}} < 1$ we find for the ground state

$$\lambda = \xi^2 \left[ -\frac{2}{a_0^2} - 4 + \frac{a_0^2}{12 a_{\text{ho}}^4} + O\left(a_0^6\right) \right]. \tag{S.32}$$

When we insert this expression into Eq. (S.26), we obtain

$$-\frac{\hbar^2}{2m} \frac{\partial^2 U_{n_\xi 0}(\xi)}{\partial \xi^2} + \left[ -\frac{\hbar^2}{2m} \frac{1}{4\xi^2} + \frac{m\omega^2}{2}(\xi - \xi_0)^2 + \frac{\hbar^2}{m}\left( -\frac{1}{a_0^2} + \frac{a_0^2}{24 a_{\text{ho}}^4} \right) \right] U_{n_\xi 0}(\xi) = E_{n_\xi 0} U_{n_\xi 0}(\xi). \tag{S.33}$$

For large $\xi_0$ the $\xi^{-2}$ term may be neglected, such that only the shifted harmonic potential remains. Hence we find the approximated hyperradial spectrum

$$E_{n_\xi 0} \approx \hbar \omega \left( n_\xi + \frac{1}{2} \right) + \frac{\hbar^2}{m} \left( -\frac{1}{a_0^2} + \frac{a_0^2}{24 a_{\text{ho}}^4} \right) + O\left( \frac{1}{\xi_0^2} \right). \tag{S.34}$$

This corresponds to the spectrum of a one-dimensional oscillator that is shifted by the binding energy.



\end{document}